\documentclass[12pt,fleqn]{article}
\RequirePackage{graphicx}
\usepackage[a4paper]{geometry}
\usepackage{multicol}
\renewcommand{\normalsize}{\fontsize{9}{11}\selectfont}
\renewcommand{\large}{\fontsize{12}{10}\selectfont}
\renewcommand{\footnotesize}{\fontsize{9}{9}\selectfont}
\title{\large
Disorder-induced neutral solitons in degenerate ground state polymers}
\author{\normalsize
Marc~Thilo~Figge\footnote{\footnotesize e-mail: figge@phys.rug.nl}, 
Maxim~V.~Mostovoy, and Jasper~Knoester\\
\normalsize
Institute for Theoretical Physics and Materials Science Center\\
\normalsize
University of Groningen, Nijenborgh 4, 9747 AG Groningen, The Netherlands}
\date{}

\begin{document}
\null
\vspace{2.0cm}
\begin{center}
{\large Disorder-induced neutral solitons in degenerate ground state polymers}\\[0.5cm]
{\normalsize Marc~Thilo~Figge\footnote{\footnotesize e-mail: figge@phys.rug.nl}, 
Maxim~V.~Mostovoy, and Jasper~Knoester\\
\normalsize
Institute for Theoretical Physics and Materials Science Center\\
\normalsize
University of Groningen, Nijenborgh 4, 9747 AG Groningen, The Netherlands}
\vspace{0.5cm}
\end{center}
%
\hrule\vspace{0.3cm}
\noindent {\bf Abstract\medskip}

We study the effects of weak off-diagonal disorder on $\pi$-conjugated 
polymers with a doubly degenerate ground-state. 
We find that disorder induces a finite density of neutral solitons in 
the lattice dimerization of a polymer chain.
Interchain interactions result in a linear potential between the solitons 
and, if sufficiently strong, bind them into pairs resulting in an
exponential suppression of the soliton density.
As neutral solitons carry spin $1/2$, they contribute to the polymer's magnetic 
properties. We calculate the magnetic susceptibility and suggest measurements 
of the magnetic susceptibility in {\it trans}-polyacetylene at low temperatures.\\

\vspace{0.3cm}\hrule
\begin{multicols}{2}
\noindent {\bf 1. Introduction\medskip}

Conjugated polymers belong to a large class of quasi-one-dimensional Peierls
materials, in which the lattice distorts due to its interaction
with itinerant electrons.  Much attention has been
devoted to {\it trans}-polyacetylene, which has a doubly degenerate
ground state.  The degeneracy allows for topological excitations,
solitons, which are kinks in the lattice dimerization accompanied
by a local distortion of the electron density. 
Solitons survive the presence of on-site Coulomb repulsion.
Although solitons are usually associated with excited states, we show
in this contribution that off-diagonal disorder may induce a finite
density of kinks in the ground state of {\it trans}-polyacetylene. 
We discuss the effects on the magnetic response.\\

\noindent {\bf 2. Mapping on the Random-Field Ising Model\medskip}

We describe the polymer chains by a Peierls-Hubbard model, which accounts for
both a static electron-lattice interaction and electron-electron interactions.
The lattice dimerization at position $n$ of the chain 
is given by:
\begin{equation}
\Delta(n) = \Delta_{lat}(n) + \eta(n) .
\end{equation}
Here, $\Delta_{lat}(n)$ describes the Peierls distortion and 
corresponds to the alternating part of the electron hopping amplitude along 
the chain direction.
The second term, $\eta(n)$, represents the disorder and stems from the 
small fluctuations in this amplitude.
While in the absence of disorder the ground state energy does not depend
on the sign of the dimerization $\Delta(n) = \pm \Delta_0$, the weak-disorder
correction to the energy,
\begin{equation}
\delta E\,\propto\,-\,\sum_{n}\Delta_{lat}(n)\,\eta(n)\;, 
\label{energycorr}
\end{equation}
removes this degeneracy. 
As we explained in Ref.~[1], the energy correction Eq.~(\ref{energycorr}) 
stabilizes neutral solitons in the lattice 
dimerization of a chain's minimal-energy lattice configuration.

As a consequence of the form Eq.~(\ref{energycorr}), the statistics of neutral 
solitons in a weakly disordered Peierls-Hubbard chain can be described by the 
random-field Ising model (RFIM) [2,3]:
\begin{equation}
E\{\sigma_{m}\} = \sum_{m=1}^{M} \left[
\frac{\mu}{2} (1-\sigma_{m}\sigma_{m+1})
-h_{m}\sigma_{m}-B\sigma_{m}\right] .
\label{rfim}
\end{equation}
The Ising variable $\sigma_m=\pm 1$ corresponds to the two possible values of the
lattice dimerization $\Delta_{lat} = \pm \Delta_0$ between neighboring kinks.
Thus, the first term in Eq.~(\ref{rfim}) is the energy cost related to the
occurence of kinks, where $\mu$ is the kink creation energy.
As the RFIM Eq.(\ref{rfim}) is an effective model, obtained by
integrating out small lattice fluctuations, $\mu$ weakly depends on temperature 
and is renormalized by electron-electron interactions [2].
Comparing Eq.~(\ref{rfim}) to Eq.~(\ref{energycorr}) reveals that the role of 
the disorder is taken over by the random ``magnetic'' field $h_m$, which we assume 
to have a Gaussian distribution with zero mean and correlator 
$\langle h_m h_n \rangle = \epsilon \delta_{m,n}$ with the
disorder strength $\epsilon\ll \mu^2$.

Finally, the third term in Eq.(\ref{rfim}) describes the interchain
interactions, which tend to establish a coherence between the
phases of the order parameter on different chains.  These
interactions are taken into account in the chain mean-field
approximation, so that the homogeneous ``magnetic'' field $B$ is
proportional to the average order parameter: $B=W 
\langle\langle\sigma\rangle\rangle$.
Here, the double brackets denote both the thermal and the
random-field average. The validity of the chain mean-field
approximation requires the interchain interaction energy to be
sufficiently weak, $W \ll \mu$, which holds in quasi-one-dimensional
materials such as conjugated polymers.

\begin{center}
\includegraphics[width=8cm]{phase.epsi}
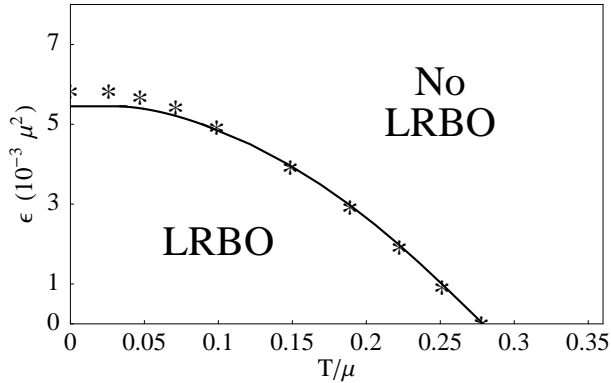
Fig.~1. Phase diagram of the RFIM.
\end{center}

We have analytically solved the model Eq.~(\ref{rfim}) 
by noticing that the disorder averaged free energy of its continuum 
version has the form of a matrix element of the Green function 
describing the relaxation of a spin $1/2$ in a time-dependent 
magnetic field.
The coordinate along the chain plays the role of the (imaginary) 
time in which the relaxation takes place. The Green function was 
found by solving the corresponding Fokker-Planck equation [2,3].

For a single chain ($B\propto W=0$) the ground state density of 
solitons is then found to be proportional to the disorder strength, 
\begin{equation}
n_s\;=\;\epsilon\,/\,\mu^2\;. 
\end{equation}
\noindent The solitons occur randomly positioned along the chain and destroy
the long-range order in the lattice dimerization at any temperature 
[2-4].
This turns out to be the same for interacting chains ($W \neq 0$), as 
long as the disorder strength is larger than the critical value 
$\epsilon_c=2W\mu/3$.
However, for $\epsilon<\epsilon_c$, the long-range bond order (LRBO) 
in the lattice dimerization is re-established as solitons are bound 
into pairs by the interchain interactions (soliton confinement).
As a consequence, the density of neutral soliton pairs $n_p$ is 
found to be exponentially suppressed at low temperatures:
\begin{equation}
n_p\;=\;2\,\frac{W^2}{\epsilon}\,\exp\left(-2\frac{W\mu}{\epsilon}\right)\;.
\end{equation}
In Fig.~1 we plot the phase diagram for the existence of LRBO in the 
RFIM Eq.~(\ref{rfim}) using parameters typical for {\it trans}-polyacetylene.
The result of a numerical calculation (stars) and our analytical result 
(solid line) are seen to be in excellent agreement.\\

\noindent {\bf 3. Magnetic Susceptibility in the Ordered Phase\medskip}

As neutral solitons carry a spin $1/2$, they contribute to the polymer's
magnetic properties.
For $\epsilon\ll\epsilon_c$, solitons only occur in isolated pairs.
Within the RFIM Eq.~(\ref{rfim}), we have analytically calculated the
pair size distribution $p(R)$, which turns out to be sharply peaked 
around a typical pair size $R=R^{\ast}$.
As the antiferromagnetic exchange between the neutral solitons of a pair
is a known (exponentially decaying) function $J(R)$ of the pair size [6],
knowledge of $p(R)$ allows us to calculate the magnetic susceptibility due
to the disorder-induced solitons.

At temperatures $T$ larger than the typical exchange value $J^{\ast}=J(R^{\ast})$, we 
find the magnetic susceptibility to obey the Curie law, $\chi(T)\propto 1/T$.
At temperatures $T\ll J(R^{\ast})$, however, most of the spin pairs are in the
singlet state, and the magnetic susceptibility (up to logarithmic corrections) reads [5]:
\begin{equation}
\chi(T)\;\propto\;\left(\frac{1}{T}\right)^{1-\alpha}\;
\label{suslowt}
\end{equation}
with $\alpha\propto W^2/\epsilon$. Thus, the interplay between interchain interactions
$(W)$ and disorder $(\epsilon)$ determines the low-temperature behaviour of the
magnetic susceptibility.\\

\noindent {\bf 4. Conclusions\medskip}

Off-diagonal disorder induces neutral solitons in the ground state of
{\it trans}-polyacetylene.
These solitons contribute to the magnetic susceptibility.
At low temperatures their magnetic response shows deviations from Curie
behavior, and the thermal behavior of the susceptibility strongly depends
on disorder and interchain interactions [Eq.~(\ref{suslowt})].
Our theory may explain the observed deviation from Curie behavior in 
Durham polyacetylene below $30 K$ [7]. 
To get better insight into the parameter $\alpha$, experiments should
be extended to temperatures of the order of $1 K$.\\

\noindent {\bf Acknowledgments\medskip}

This work is supported by the Stichting Fundamenteel Onderzoek der Materie (FOM).\\

\noindent {\bf References\medskip}

\noindent [1] M. Mostovoy, M. T. Figge, and J. Knoester, 
Europhys. Lett. 38 (1997) 687.

\noindent [2] M. Mostovoy, M. T. Figge, and J. Knoester, 
Phys. Rev. B 57 (1998) 2861.

\noindent [3] M. T. Figge, M. Mostovoy, and J. Knoester, 
Phys. Rev. B 58 (5) (1998); cond-mat/9803385.

\noindent [4] Y. Imry and S. Ma, Phys. Rev. Lett. 35 (1975) 
1399.

\noindent [5] M. T. Figge, M. Mostovoy, and J. Knoester,
{\it submitted to Phys. Rev. B}.

\noindent [6] Y.R. Lin-Liu and K. Maki,
Phys. Rev. 22 (1980) 5754.

\noindent [7] P.J.S. Foot, N.C. Billingham, and P.D. Calvert,
Synth. Met. 16 (1986) 265.
\end{multicols}
\end{document}